\documentclass[epj,final]{svjour}[30-12-2013]
\usepackage{amssymb}
\usepackage{amsmath}
\usepackage{graphicx}
\usepackage{hyperref}
\usepackage{epsfig}
\usepackage{color}
\usepackage{dcolumn}
\usepackage{bm}
\usepackage{epstopdf}

\begin{document}

\title{Spectral flow of trimer states of two heavy impurities and one light condensed boson}

\author{N.~T. Zinner}
\institute{Department of Physics and Astronomy, Aarhus University, Aarhus C, DK-8000}

\date{\today}

\abstract{
The spectral flow of three-body (trimer) states consisting of two heavy (impurity) particles
sitting in a condensate of light bosons is considered. Assuming that the condensate is 
weakly interacting and that an impurity and a boson have a resonant zero-range two-body 
interaction, we use the 
Born-Oppenheimer approximation to determine the effective three-body potential. We solve
the resulting Schr{\"o}dinger equation numerically and determine the trimer binding 
energies as a function of the coherence length of the light bosonic condensate particles.
The binding energy is found to be suppressed by the presence of the condensate when the 
energy scale corresponding to the coherence length becomes of order the trimer binding 
energy in the absence of the condensate. We find that the Efimov scaling property is 
reflected in the critical values of the condensate coherence length at which the trimers
are pushed into the continuum.
\PACS{
{03.65.Ge}{Solutions of wave equations: bound states } \and
{03.75.Hh}{Static properties of condensates; thermodynamical, statistical, and structural properties} \and
{67.85.-d}{Ultracold gases, trapped gases}}
}

\maketitle

\section{Introduction}
The study of few-body quantum mechanical bound states formed via strong short-range interactions
has seen a tremendous boost after the observation of three-body states in cold atomic
gases \cite{kraemer2006,ferlaino2010}. This is tied to a famous prediction by Vitaly Efimov 
that an infinitude of three-body bound states of identical bosons occurs at the point where
the two-body subsystems have a bound state exactly at zero energy \cite{efimov1970}. These
curious states have several interesting properties that are all in some way connected to 
the exponential factor $e^{\pi/s_0}$ where $s_0$ is called the scaling parameter. For instance,
the three-body binding energies will scale with a factor of $e^{2\pi/s_0}\sim 515$, becoming 
weaker by a huge factor for each subsequent state in the spectrum. Another key feature of these
so-called Efimov states is that away from the resonance where the two-body binding energy 
vanishes, they display so-called borromean binding where no two-body subsystem is bound yet the three-body
states remains bound (see Ref.~\cite{baas2012} for a discussion of this phenomenon in the context of 
three or more particles). 

The severe reduction in binding energy for subsequent states in the three-body spectrum 
strongly limits the number of states that one can expect to observe in experiments. 
However, for three-body systems with particles of different mass, this factor can 
be reduced considerably \cite{jensen2004,bra2006} and more states could possibly be observed when using a
mixture of different atomic species \cite{ferlaino2010}. Meanwhile, 
many ultracold atom 
experiments probe many-body systems which means that atomic few-body
states should be considered as embedded in an environment. This has little influence on the
few-body physics at 
very low density where theory and experiments match well, but the ability to push into 
higher density regimes is there and it is desirable in order to connect these studies to
systems in other branches of physics such as condensed-matter and solid-state physics. 

The question of what happens to few-body states when the Pauli principle influences one or
several of the particles that make up the bound state has been discused in a number of recent
papers \cite{nishida2009,macneill2011,nygaard2011,song2011,niemann2012,endo2013}. Similarly,
a number of works have addressed strongly interacting Bose gases at the border between
the few-body and many-body perspective \cite{borzov2012,zhou2012,mashayekhi2013,jiang2013,komnik2013,rath2013}.
Also, some cases of one or more impurities in a condensate have been considered in the 
recent past \cite{bijlsma2000,viverit2000,cucchietti2006,kalas2006,tempere2009,casteels2011,santamore2011}.
In the current paper, we consider two heavy impurities that interact with a light bosonic 
particle which produces a small scale factor and thus should be a case where a number of 
three-body bound states may be observed. The effect of the many-body environment comes
about since we assume that the light boson is part of a Bose condensate. This problem
is reminiscient of electron-electron interactions mediated by phonons with the latter
acting as the light bosonic particles that are condensed. However, we do take into
account that the dispersion of the light boson is only linear at low momenta and then
quadratic at higher momenta. The physics bears resemblence to a bosonic bath version of the 
Kondo problem studied in the condensed-matter community 
\cite{affleck1995,smith1999,sengupta2000,vojta2000,zarand2002,florens2006}. 

The current study 
extends our recent paper \cite{zinner2013} that suggests a direct and dramatic effect of the 
condensate environment on the Efimov three-body spectrum. Assuming that the condensate
is in the weak coupling limit with a small and positive scattering length, $a_B>0$,
we found in Ref.~\cite{zinner2013}, that the 
condensate coherence length, $\xi=1/\sqrt{8\pi n_0 a_B}$ with $n_0$ the condensate gas density,
may interfere with the three-body potential and reduce the binding until the states
are eventually destroyed. Here we study this issue in a more quantative manner by 
numerical calculation of the spectrum as a function of $\xi$. This confirms and 
expands on the ideas presented in Ref.~\cite{zinner2013}. Furthermore,
we find that the spectral flow of the three-body binding energies with $\xi$ 
displays an Efimov scaling similar to what has been discussed for fermions recently
\cite{macneill2011,nygaard2011,niemann2012}.
We also provide some appendices with
a discussion of the technical details used here and in Ref.~\cite{zinner2013}.

\section{Model}\label{theory}
Our basic setup consists of a quantum gas of bosons with mass $m$ and two embedded 
impurities with mass $M$ that we assume are much heavier than the bosons, i.e. $m\ll M$
The interaction is of the density-density type and in second quantization the 
Hamiltonian becomes
\begin{eqnarray}
&H=\sum_{\bm k} \epsilon_I(\bm k)c_{\bm k}^{\dagger}c_{\bm k}+\sum_{\bm k}\epsilon_B(\bm k)b_{\bm k}^{\dagger}b_{\bm k}&\nonumber\\
&+U_{B}\sum_{\bm q}n_B({\bm q})n_B({-\bm q})+U_{IB}\sum_{\bm q}n_B({\bm q})n_I({-\bm q}),&
\end{eqnarray}
where $c_{\bm k}$ are the impurity operators and $b_{\bm k}$ the boson operators. The interactions are
assumed to be zero range with $n_I(\bm q)=\sum_{\bm k}c_{\bm k+\bm q}^{\dagger}c_{\bm k}$ and
$n_B(\bm q)=\sum_{\bm k}b_{\bm k+\bm q}^{\dagger}b_{\bm k}$. 
The single-particle dispersions are $\epsilon_I(\bm k)={\bm k}^2/2M$ and $\epsilon_B(\bm k)={\bm k}^2/2m$
respectively. The impurity-boson coupling $U_{IB}$ will be addressed below.
When $a_B$ is small we may use the weak coupling limit result 
$U_B=4\pi\hbar^2a_B/m$. We use this expression throughout this paper and thus assume 
that the bosons are weakly interacting. In terms of the coherence length, $\xi$, this means
that we are assuming that it is large. Below we will also push into the regime where $\xi$ 
becomes of the same order as the impurity-boson scattering length, $a$. In the regime
where $\xi\leq |a|$ our method is not expected to hold quantitatively but should reveal
the qualitative features of the setup.

In a weakly-interacting Bose gas, Bogoliubov theory applies to 
the light bosonic particles \cite{fetter1971}. We therefore transform to quasi-particles, $\gamma$
and $\gamma^\dagger$, via
\begin{eqnarray}
&b_{\bm k}^{\dagger}=u_{\bm k}\gamma_{\bm k}^{\dagger}+v_{\bm k}\gamma_{-\bm k}&\\
&b_{-\bm k}=v_{\bm k}\gamma_{\bm k}^{\dagger}+u_{\bm k}\gamma_{-\bm k}&,
\end{eqnarray}
where $u,v$ are assumed real and symmetric in $\bm k$. The Hamiltonian 
is now
\begin{eqnarray}
&H=\sum_{\bm k} \epsilon_I(\bm k)c_{\bm k}^{\dagger}c_{\bm k}+\sum_{\bm k\neq 0}E_B(\bm k)\gamma_{\bm k}^{\dagger}\gamma_{\bm k}&\nonumber\\
&+U_{IB}\sum_{\bm q}n_B({\bm q})n_I({-\bm q}),&
\end{eqnarray}
with $E_B(\bm k)=\sqrt{U_B n_0{\hbar^2\bm k}^2/m_B+({\hbar^2\bm k}^2/2m_B)^2}$
where $n_0$ is the condensate density. Constant terms have been dropped as they are 
are not relevant here. 
The boson density operator can now be expressed in terms of the quasiparticle operators. The $\bm q=0$ terms are
\begin{eqnarray}
n_B(\bm q)=n_0\delta_{\bm q 0}+\sqrt{n_0}\left(b_{\bm q}^{\dagger}+b_{-\bm q}\right)+\sum_{\bm k\neq 0}b_{\bm k+\bm q}^{\dagger}b_{\bm k}.
\end{eqnarray} 
Transforming to $\gamma$ operators one obtains
\begin{eqnarray}
&n_B(\bm q)=n_0\delta_{\bm q 0}+\sqrt{n_0}\left(u_{\bm q}+v_{\bm q}\right)\left(\gamma_{\bm q}^{\dagger}+\gamma_{-\bm q}\right)&\\
&+\sum_{\bm k\neq 0}\left[u_{\bm k+\bm q}v_{\bm k}\gamma_{\bm k+\bm q}^{\dagger}\gamma_{-\bm k}^{\dagger}
+u_{\bm k+\bm q}u_{\bm k}\gamma_{\bm k+\bm q}^{\dagger}\gamma_{\bm k}^{}\right.&\\
&\left.+v_{\bm k+\bm q}v_{\bm k}\gamma_{-\bm k-\bm q}^{}\gamma_{-\bm k}^{\dagger}
+v_{\bm k+\bm q}u_{\bm k}\gamma_{-\bm k-\bm q}^{}\gamma_{\bm k}^{}\right].&
\end{eqnarray}
The number of terms is reducible by using $u\to 1$ and $v\to 0$ in the 
weakly interacting limit. 
This eliminates the $\gamma\gamma$ and $\gamma^\dagger\gamma^\dagger$ pieces. 
Also, the first term will yield a constant proportional to the impurity density when inserted
in the Hamiltonian so we drop this also. The most problematic term is the one linear in 
$\gamma$ and $\gamma^\dagger$ with a prefactor of $\sqrt{n_0}$. Here we will assume that the
condensate density is small, which in the weak-coupling limit, is the same as 
assuming that $n_B$ is small. The impurity-boson interaction term now reduces to
\begin{eqnarray}
U_{IB}\sum_{\bm q\bm k\bm k'}c_{\bm k-\bm q}^{\dagger}c_{\bm k}\gamma^{\dagger}_{\bm k'+\bm q}\gamma_{\bm k'},
\end{eqnarray}
which is a standard interaction term for a zero-range interaction of strength $U_{IB}$.
The effective Hamiltonian is then
\begin{eqnarray}
&H=\sum_{\bm k} \epsilon_I(\bm k)c_{\bm k}^{\dagger}c_{\bm k}+\sum_{\bm k\neq 0}E(\bm k)\gamma_{\bm k}^{\dagger}\gamma_{\bm k}&\nonumber\\
&+U_{IB}\sum_{\bm q\bm k\bm k'}c_{\bm k-\bm q}^{\dagger}c_{\bm k}\gamma^{\dagger}_{\bm k'+\bm q}\gamma_{\bm k'},&
\end{eqnarray}
which corresponds to impurity particles with dispersion $\epsilon_I(\bm k)$ interacting with Bose gas particles with 
dispersion $E(\bm k)$ through a contact interaction with strength $U_{IB}$. In the case where $E(\bm k)$ is 
linear in $\bm k$, this corresponds to a system of (heavy) electrons interacting with phonons through a non-dispersive
zero-range intearction. 

Before we proceed to solve the three-body problem of two heavy impurities and one
light boson in a condensate using this effective Hamiltonian, a couple of remarks
are in order. The states we discuss here are {\it not} the absolute ground state
of the system, but rather few-body resonances similar to the ones that can be 
studied in cold atomic gases \cite{ferlaino2010}. The true ground state in the
current model is more likely a bound state of both impurities and all the bosons
in the condensate. However, such large cluster are not formed on the typical 
experimental timescale in cold atomic gases. An alternative bound state to 
consider would be one with two light bosons and one impurity. This type of 
system has a very large Efimov scale factor $e^{\pi/s_0}$ \cite{jensen2004,bra2006} and observation 
of Efimov states is therefore not favoured. In constrast, two heavy impurities
and one light particle has a small scale factor and the spectrum of Efimov
states is expected to be dense (see Appendix~\ref{appC} for a discussion of 
this feature using the Born-Oppenheimer approximation).

\subsection{Born-Oppenheimer Approximation} 
Since we model the impurity-boson interaction by a zero-range two-body
potential, we have 
\begin{equation}\label{impot}
V(\bm r)=V_0\left[\delta(\bm r-\bm R/2)+\delta(\bm r+\bm R/2)\right],
\end{equation}
where the impurities are located at $\pm \bm R/2$. Regularization is 
necessary and we discuss this momentarily. We will now use the 
Born-Oppenhimer approximation. Here the light particle motion is 
solved while assuming that $\bm R$ is unaltered. This is 
done for all values of $\bm R$, and in turn one obtains an effective 
Schr{\"o}dinger equation for the heavy particles which depends on $\bm R$.
This implies that we consider $\bm R$ an adiabatic variable that 
changes much slower than the position of the light particle.

The Schr{\"o}dinger
equation for the light particle of mass $m$ in the potential of Eq.~\eqref{impot} is
\begin{equation}
H\phi=E_R\phi,
\end{equation}
where $E_R$ is the energy and $\phi$ the wave function. Delta functions are 
conveniently handled in momentum space and we thus transform our equation 
into
\begin{equation}
E({\bm k})\phi(\bm k)+\frac{1}{(2\pi)^3}\int d^3k' \phi(\bm k')V(\bm k-\bm k')=E_R\phi(\bm k).
\end{equation}
For the case at hand we have
$V(\bm q)=2V_0 \cos\left(\bm q\cdot \frac{\bm R}{2}\right)$ and in turn
\begin{align}
%(E({\bm k})-E_R)
\phi(\bm k)=\frac{-2V_0}{(2\pi)^3}\frac{1}{E({\bm k})-E_R}\int d^3k' \cos\left(\frac{(\bm k-\bm k')\cdot \bm R}{2}\right)\phi(\bm k').
\end{align}
One may now multiply by $\cos(\bm k\cdot \bm R/2)$ and integrate both sides with respect to ${\bm k}$. Using the assumptions that $\phi$ is even in ${\bm k}$ (solutions with $s$-wave symmetry),
a couple of trigonometric manipulations leads us to the equation
\begin{equation}\label{nonreg}
1=-\frac{V_0}{(2\pi)^3}\int d^3k \frac{1}{E({\bm k})-E_R}-\frac{V_0}{(2\pi)^3}\int d^3k \frac{\cos\left(\bm k\cdot\bm R\right)}{E({\bm k})-E_R}.
\end{equation}

It is now necessary to face the issue of regularization of the interaction, which implies that the 
parameter $V_0$ be substituted for the physical impurity-boson scattering length, $a$, while at the 
same time absorbing the divergence that appears in the first integral on the right-hand side 
of Eq.~\eqref{nonreg}. This can be done elegantly using pseudopotentials \cite{tan2008,valiente2012}, 
although here we will use the more traditional Lippmann-Schwinger equation approach.
The scattering equation for the impurity-boson problem with a zero-range two-body interaction is
\begin{equation}\label{lipp}
\frac{1}{V_0}=\frac{\mu}{2\pi a\hbar^2}-\frac{1}{(2\pi)^3}\int d^3k\frac{1}{\epsilon^{\mu}_{\bm k}},
\end{equation}
with the reduced mass $\mu=mM/(m+M)$ and the reduced energy $\epsilon^{\mu}_{\bm k}=\hbar^2k^2/2\mu$. 
Since we work in the limit $m\ll M$, we take $\epsilon^{\mu}_{\bm k}=\epsilon_{\bm k}$ and $\mu=m$. 
Here we use the 
bare single-particle dispersion, $\hbar^2{\bm k}^2/2m$, for the bosons and {\it not} $E(\bm k)$. 
We do so because the impurity-boson scattering length, $a$, is defined in vacuum and the scattering 
equation must thus be the vacuum version as well.
Combining Eq.~\eqref{lipp} and Eq.~\eqref{nonreg} we obtain the regularized momentum space
Schr{\"o}dinger equation
\begin{align}
&\frac{m}{2\pi a\hbar^2}=-\frac{1}{(2\pi)^3}\int d^3k \left[\frac{1}{E({\bm k})-E_R}-\frac{1}{\hbar^2{\bm k}^2/2m}\right]&\nonumber\\
&-\frac{1}{(2\pi)^3}\int d^3k \frac{\cos\left(\bm k\cdot\bm R\right)}{E({\bm k})-E_R},&
\end{align}
which can be written
\begin{align}
\label{central}
&\frac{R}{a}=-\frac{2}{\pi}\alpha R \int_{0}^{\infty} dx \left[\frac{x^2}{[x^4+A^2x^2]^{1/2}+1}-1\right]&\nonumber\\
&-\frac{2}{\pi}\int_{0}^{\infty} dx \frac{x \sin(\alpha R x)}{[x^4+A^2x^2]^{1/2}+1},&
\end{align}
when we define the dimensionless quantities $\alpha^2=-2mE_R/\hbar^2$ and $A=1/(\alpha\xi)$. The 
two integrals in Eq.~\eqref{central} are discussed in various limits in Appendix~\ref{appA}.

For non-interacting bosons with $a_B=0$, we have $\xi=\infty$ and correspondingly $A=0$. Using the 
analytical results in Appendix~\ref{appA}, one obtains the formula
\begin{equation}\label{eig}
\alpha R=\frac{R}{a}+e^{-\alpha R}.
\end{equation}
When the impurity-boson interaction is on resonance, $|a|=\infty$, the solution is $\alpha R=x_0\sim 0.567$. 
One can obtain the solution as a power series in the variable $R/a$, and to lowest order one finds
\begin{equation}
E_R=-\frac{\hbar^2x_{0}^{2}}{2mR^2}\left[1+\frac{1}{x_{0}^{2}(1+e^{x_0})}\frac{R}{a}\right].\label{noBEC}
\end{equation}
In the limit $R\gg a$, the second term on the right-hand side of Eq.~\eqref{eig} goes to zero and
one obtains the well-known result
\begin{equation}
E_R=-\frac{\hbar^2}{2ma^2}.
\end{equation}
The physical interpretation is that the light particle becomes bound to either of the impurities 
which are effectively just static potentials in space.
The binding energy is the one appropriate for a zero-range two-body potential and the solution only exists
for $a>0$ where the impurity-boson system can bind.

\section{Three-Body Bound States}
\begin{figure}
\centering
\includegraphics[scale=0.45]{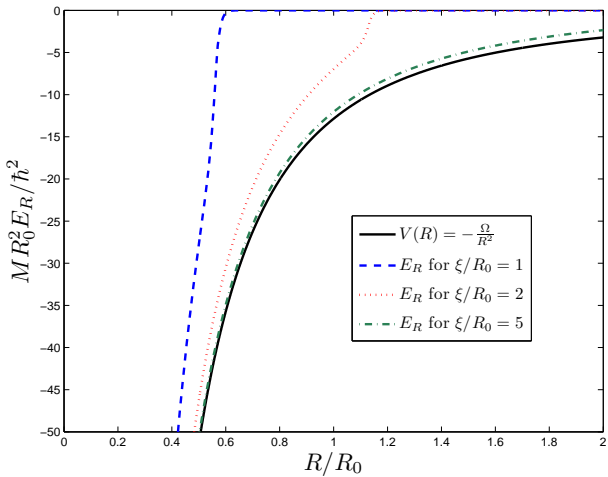}
\caption{Potential, $E_R$, between the two impurities in the Born-Oppenheimer approximation as function 
of $R/R_0$ where $R_0$ is the short-distance cut-off (see discussion in text) which is the unit of length. 
$E_R$ is shown
for $\xi/R_0=1$ (dashed line), $\xi/R_0=2$ (dotted line), and $\xi/R_0=5$ (dot-dashed line). 
For comparison, the solid line shows the potential, $V(R)=-\Omega/R^2$, when the light particle is not 
part of a condensate. In all cases shown, 
the strength of the potential is $\Omega=40x_{0}^{2}$ corresponding to a mass ratio
$M/m=80$.}
\label{potfig}
\end{figure}

The two-body bound state solutions of a single impurity and a light bosonic particle 
were discussed in details in Ref.~\cite{zinner2013} and we will not 
repeat this discussion here but rather move onto the three-body problem.
The influence of the condensate on the light particle goes through the 
dispersion, $E(\bm k)$, and then into the properties of the 
heavy-heavy-light three-body bound states via the Born-Oppenheimer
approximation. In turn, we need to determine the 
impurity-impurity potential, $E_R$, which is given implicitly
through $\alpha$ which is obtained by solving Eq.~\eqref{central}.
The first integral in Eq.~\eqref{central} is exactly solvable, see
Appendix~\ref{appA}. However, the second term in Eq.~\eqref{central}
is more difficult both analytically and numerically. 
As discussed in Ref.~\cite{zinner2013}, it can be solved using 
different approximate approaches that are discussed in more detail
in Appendix~\ref{appA}. The main message is that the presence
of a condensate influencing the light particle tends to suppress
the attractive $1/R^2$ potential that is obtained in the non-condensed
limit where $\xi\to\infty$ and $E_R \propto -1/R^2$ (see the solid 
line in Fig.~\ref{potfig}).

\subsection{Impurity-impurity Potential}
In the non-condensed limit $\xi\to\infty$, a finite value of 
$a$ would imply that the $-1/R^2$ behaviour is cut off at
$R\sim|a|$ \cite{jensen2004,bra2006}. The qualitative implication
is that whichever is smaller of $|a|$ and $\xi$ will cause 
$E_R$ to fall off to zero faster than $-1/R^2$ for $R\gtrsim \textrm{min}(|a|,\xi)$.
Here we will focus on the limit where the impurity-light boson
two-body interaction is resonant, i.e. we assume that $a\to\infty$
and thus we neglect the left-hand side of Eq.~\ref{central}. The 
modification of the $-1/R^2$ behaviour is therefore controlled 
by the value of $\xi$.

As discussed in Ref.~\cite{zinner2013}, the numerical solution is increasingly difficult to 
find for $R/\xi\to x_0$, but strongly indicates that $\alpha R$ goes to zero at a finite
value of $R/\xi$. The behaviour can be accurately fitted by the function
\begin{align}\label{potfit}
\alpha R =\frac{x_0}{2}+\frac{x_0}{2}\left[1-\left(\frac{R}{x_0\xi}\right)^2\right]^{1/2},
\end{align}
for $\alpha R>x_0/2$. Notice that there is a typo in \cite{zinner2013} where the 
first term $x_0/2$ is missing and a factor of $1/2$ is missing from the second term. Alternatively, we can approach 
the limit of small $R\sim \xi$ and $\alpha\xi$ by 
considering the limit $A^2\gg 1$ 
before doing the integrations in Eq.~\eqref{central}. The asymptotic expressions are
discussed in Appendix~\ref{appA} and given in Eqs.~\eqref{kinasymp} and \eqref{sinasymp}. In 
this limit, Eq.~\eqref{central} yields
\begin{align}\label{upper}
(\alpha\xi)^2=\frac{2}{\pi}\frac{\frac{\xi}{R}-\frac{R}{\xi}}{\frac{R}{\xi}-1}<0.
\end{align}
This is manifestly negative and implies that $E_R$ changes sign as $R\sim \xi$. We caution that
in the limit $R\to\xi$ in Eq.~\eqref{upper} the result is $-\tfrac{4}{\pi}$. In turn $A^2$ is of order 1
which is conflicting with the limit we used to derive the expression. In any case, it strongly suggests
that $E_R$ goes to zero at a finite $R$, and we will assume that this is generally the case
for any finite value of $\xi$. 

Combining the accurate fit in Eq.~\eqref{potfit} with the fact that $E_R$ must vanish beyond some 
finite $R$, we now introduce our model to describe the impurity-impurity potential.
We will assume that the fit in Eq.~\eqref{potfit} holds from 
$R=0$ to $R/\xi=x_0$, i.e. when the expression under the root in Eq.~\eqref{potfit} is non-negative.
As discussed in Ref.~\cite{zinner2013}, the presence of the condensate influences the potential
$E_R$ at large distance by making it approach zero rapidly for $R\sim \xi x_0$. 
Furthermore, we assume that as long as $E_R$ goes to zero faster than $1/R^2$ when $R\sim \xi x_0$, the functional form is not
important. We therefore use a Fermi-Dirac function that changes rapidly from 1 to 0 around
$R/\xi=x_0$. In total, the potential, $E_R$, in the spherically symmetric $s$-wave channel 
between the two impurities becomes
\begin{equation}\label{fullpot}
E_R=-\frac{\hbar^2}{2mR^2}\frac{x_{0}^{2}}{4}\left[1+\sqrt{1-(R/x_0\xi)^2}\right]^2 F[R,\xi x_0,w],
\end{equation}
where the Fermi-Dirac function is
\begin{equation}
F[x,y,z]=\frac{1}{1+e^{(x-y)/z}},
\end{equation}
and $w$ is a width parameter that we set to $w=0.01R_0$. Here $R_0$ is a short-distance cut-off
that we discuss momentarily. We have checked that small changes in $w$ 
does not affect the physics discussed in this paper. 
In the limits where $\xi\to \infty$, 
we recover the attractive $1/R^2$ potential that gives rise to the Efimov effect (see Appendix~\ref{appC}).
The resulting potentials for different values of $\xi$ are shown in Fig.~\ref{potfig} for mass ratio $M/m=80$.
We notice in the plot that already for $\xi/R_0=5$, the limiting $-1/R^2$ behaviour is approached. 

The number of three-body Efimov state expected in the absence of the condensate ($\xi=\infty$) is 
\begin{equation}
N_B\approx \frac{s_0}{\pi} \textrm{Log}\left[\frac{|a|}{R_0}\right], \label{number}
\end{equation}
where $R_0$ is a short-distance cut-off which we return to below 
(also discussed in Appendix~\ref{appC}).
For finite $\xi$ we clearly see the attractive region shrinking in Fig.~\ref{potfig} and one can
in turn replace $a$ by $\xi$ in the formula for $N_B$ as a crude estimate. 
In general, whichever is smaller of $a$ and $\xi$ sets the scale for the extend of the attraction 
and thus the number of three-body Efimov states that can be accomodated. This is the basic
physical idea presented in Ref.~\cite{zinner2013}. In the next section we proceed to discuss this
idea quantitatively using the potential in Eq.~\eqref{fullpot}.

As noted in Ref.~\cite{zinner2013}, the current model does not imply any influence of the 
condensate on the short distance behaviour of the impurity-impurity potential. 
Short distance corresponds to high momentum for which the single-particle dispersion of the 
light bosonic particles, $E(\bm k)$, goes to the free particle dispersion. The 
short distance cut-off, $R_0$, introduced above should therefore be interpreted 
and treated in the usual manner. While it is often considered merely an effective
parameter (denoted the three-body parameter), in atomic systems it appears to 
have a much more direct physical connection to the two-body inter-atomic 
potentials 
and can be related to the van der Waals length, $r_\textrm{vdW}$
\cite{berninger2011,naidon2011,chin2011,schmidt2012,wang2012,peder2012,naidon2012,wangwang2012,peder2013a,peder2013b}. 
In the current study  we assume $m\ll M$. In this case it is quite natural 
from the discussion above that $E_R$ will be regularized at short distance by
a cut-off that comes from the impurity-impurity potential. In other words, 
the regularization is given by the short-distance behaviour of the two-body
potential between the two impurities.

\begin{figure}
\centering
\includegraphics[scale=0.45]{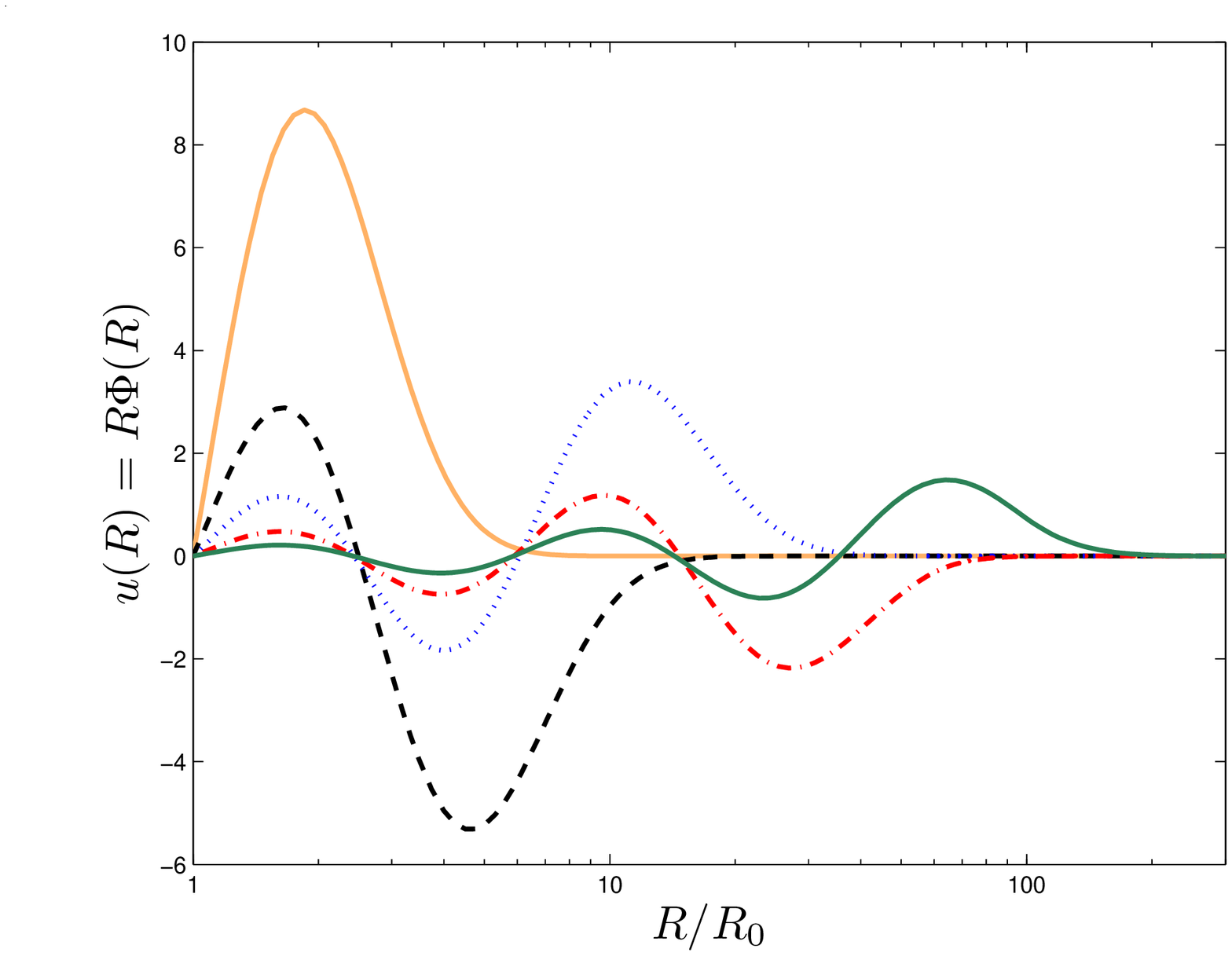}
\caption{Eigenstates, $u(R)$, of the Schr{\"o}dinger equation in Eq.~\eqref{sch} 
for $M/m=80$ and with a short-distance cut-off at $R=R_0$ as function of $R/R_0$. 
The horizontal axis has a logarithmic scale for increased visibility. 
Notice how the first node of the wave function 
coincides for the four excited states, the second node for the three highest
excited states, and the third node for the two highest excited excited states. 
This is a key feature of three-body Efimov states.}
\label{waves}
\end{figure}

\subsection{Solutions and Spectral Flow}
We now present a set of solutions for the wave functions, $\Phi(R)$, and the 
energies, $E$, of the Schr{\"o}dinger equation in the spherically 
symmetric $s$-wave channel for the potential $E_R$ in Eq.~\eqref{fullpot}, i.e.
\begin{equation}\label{sch}
\left[-\frac{\hbar^2}{M}\frac{d^2}{dR^2}+E_R\right]u(R)=Eu(R),
\end{equation}
where $u(R)=R\Phi(R)$. We take a large $M/m=80$ when computing 
the scale factor, $s_0$. The
large ratio of $M/m$ is employed to have several states to study as a function 
of $\xi$. For other mass ratios we find similar behavior.
The scale factor can now be 
computed using the formulas in Appendix~\ref{appC} and we obtain $s_0=3.55$
or $e^{\pi/s_0}=2.42$. We also need to specify the short distance cut-off
and here we take $R_0$. Since we assume that $|a|\to\infty$, the 
limit where also $\xi\to\infty$ in principle has an infinite number
of three-body bound states due to the Efimov effect. The numerical 
precision of course limits the number that can actually be found in a
given calculation and we find $N_B=8$ for large $\xi$ as we 
discuss below.

In Fig.~\ref{waves} we show $u(R)$ for the five lowest bound states obtained by
numerical solution of the Schr{\"o}dinger equation for large $\xi$. This was done by
direct integration of the inner part of the potential and then a 
comparison to the exact bound state exponential tail outside the 
potential (with the boundary defined so that the potential can be 
neglected, i.e. for $R>\xi x_0$, see Fig.~\ref{potfig}). The five bound 
states in Fig.~\ref{waves} nicely follow the rules of adding an additional 
node to the wave function as one moves up in the spectrum. Notice
the first zero of the four excited states which is at the same 
value of $R/R_0$ (and likewise for the second node of the three highest
states, and third node for the two highest states). This is a 
clear manifestation of the scale invariance of the wave function 
that is the essence of the Efimov effect (see Appendix~\ref{appC}).
Armed with the knowledge that our numerical procedure produces an 
Efimov spectrum with a number of states, we proceed to consider 
finite values of $\xi$.

\begin{figure}
\centering
\includegraphics[scale=0.45]{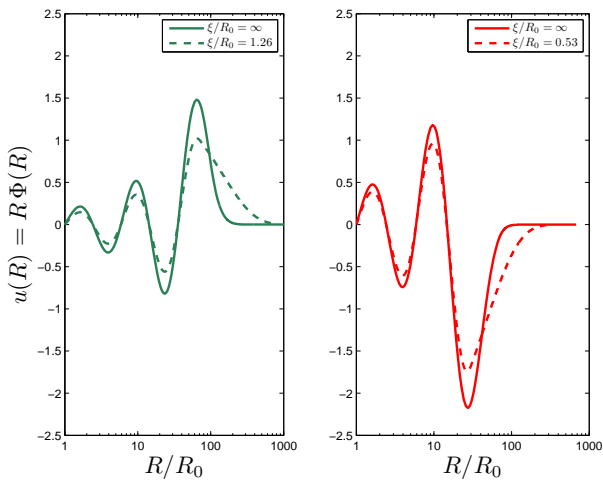}
\caption{Eigenstates, $u(R)$, as in Fig.~\ref{waves} but for both $\xi/R_0=\infty$ and
for finite $\xi$. Left panel: The fourth excited state for $\xi/R_0=\infty$ (solid line) and $\xi/R_0=1.26$
(dashed line). Right panel: The third excited state for $\xi/R_0=\infty$ (solid line) and $\xi/R_0=0.53$
(dashed line). Notice the transfer of wave function amplitude to larger $R/R_0$ when $\xi$ is finite.
This shows the modifications caused by the condensate at large distance.}
\label{modwaves}
\end{figure}

The modifications to the wave functions are illustrated in Fig.~\ref{modwaves} for the fourth
excited state (left panel) and third excited state (right panel). In both cases we see that for
finite $\xi/R_0$, the wave function is mainly modified at large distance. In fact, the wave function 
amplitude seems to spread to large distance in comparison to the $\xi/R_0=\infty$ case. This 
observation is perfectly consistent with the discussion above and the subsequent expectation that 
the presence of a condensate affecting the light particle should influence the large distance 
bahaviour. This flow of probability to the outer regions can be understood from the 
potential form in Eq.~\eqref{fullpot} or more precisely from Eq.~\eqref{potfit}.
The potential goes rapidly to zero at $R/\xi\sim x_0$ as shown in Fig.~\ref{potfig} and the strength
of the attractive $1/R^2$ decreases with decreasing $\xi$. This means that the wave function 
can more easily leak out into the region beyond $R/\xi=x_0$ as we see in Fig.~\ref{modwaves}.
An increasing amplitude in the region of space where the potential is negligible implies that 
the binding energy must subsequently decrease since the potential is purely attractive. We will
now quantify this feature and demonstrate the spectral flow with $\xi$ and its connection to 
Efimov scaling.

\begin{figure}
\centering
\includegraphics[scale=0.45]{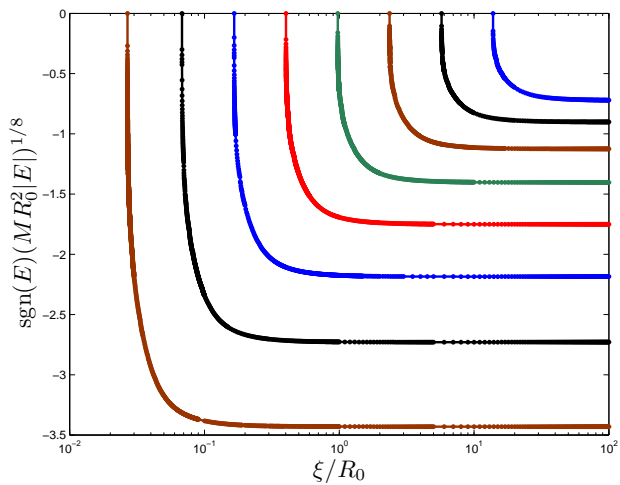}
\caption{Spectral flow as a function of $\xi/R_0$ of the energy for the four excited states shown in Fig.~\ref{waves}. 
The vertical axis has been scale by the power $1/8$, while the horizontal axis is logarithmic. This has been done to 
make the connection to the Efimov scaling factor, $e^{\pi/s_0}$, clearly visible in the plot.
The turning points at which the binding energy starts to decrease rapidly is roughly at the point where 
$\hbar^2/M\xi^2$ is equal to the binding energy 
of the state for $\xi=\infty$.}
\label{flow}
\end{figure}

The variation of the binding energy is shown in 
Fig.~\ref{flow} as a function of $\xi/R_0$. The plot has been rescaled on the vertical axis by a power 
of $1/8$ (see axis label), while the horizontal axis is logarithmic. This helps visualize the presence
of Efimov scaling as $\xi$ changes. Indeed in the plot we clearly see the self-similar behaviour of 
the flow. The weakest bound state is influenced by the condensate at the largest $\xi$, and subsequently
has the largest critical value, $\xi_c$, at which its binding goes to zero. Then the next deeper bound
state repeats the same flow as $\xi$ decreases still further and so on. The jump 
in the energy seen in Fig.~\ref{flow} happens when the energy of the three-body state in 
the limit $\xi/R_0=\infty$ (absence of condensate effects) is roughly equal to $E_\xi=\hbar^2/M\xi^2$. This 
means that we have a sort of 'window of opportunity' for a given $\xi$ value within which modifications
to three-body states can be expected. By comparison to the spectral flow of Efimov
three-body states in the presence of a Fermi sea \cite{macneill2011,nygaard2011}, we see that the 
Fermi energy scale, $E_F$, is replaced by $E_\xi$. Apart from this, the two flows of the three-body 
binding energies are very similar. Note, however, that this mapping of similarities does not hold
for the two-body impurity-boson system as discussed in Ref.~\cite{zinner2013} where light bosons 
can instead be mapped to a problem with heavy fermions (at least within the Born-Oppenheimer approach).
 
The signatures of Efimov scaling are clearly seen in Fig.~\ref{flow}. To further quantify this finding
we provide an overview of the different scaling factors of adjacent states in Tab.~\ref{tab}. The
table contains ratios of binding energies and also ratios of critical coherence lengths, $\xi_c$, where
the binding energies vanish. The scaling factors obtained numerically from the binding energies are close
to the analytically calculated ratios and only for the 
lowest and highest states, do we see deviations. This is not unexpected. In the case of 
the lowest state, the binding energy is very large and the short-distance part of the potential 
plays a role through the cut-off $R_0$. For the highest excited state that has
substantial amplitude at larger distances, the deviation comes from sensitivity to 
the fact that the potential is modified from the attractive inverse square form at 
large distance but nevertheless this remains a small effect.
The story is similar for the ratio 
of critical coherence lengths as seen in Tab.~\ref{tab}.

\begin{table}
\centering
\begin{tabular}{ | c | c | c| c| c | }
\hline
 $n $ & $1$ & $2$ & $3$ & $4$  \\
\hline
 $E_{0}^{n-1}/E_{0}^{n} $ & $6.23$ & $5.92$ & $5.88$ & $5.89$\\
 $\xi_{c}^{n-1}/\xi_{c}^{n}$ & $2.54$ & $2.43$ & $2.42$ & $2.42$ \\
\hline
 $n $ & $5$ & $6$ & $7$ & $\infty$  \\
\hline
 $E_{0}^{n-1}/E_{0}^{n} $ & $5.85$ & $5.89$ & $6.01$ & $5.87$\\
 $\xi_{c}^{n-1}/\xi_{c}^{n}$ & $2.42$ & $2.43$ & $2.41$ & $2.42$ \\
\hline
\end{tabular}
\caption{Ratios of energies and of critical coherence lengths, $\xi_c$, at which the 
binding energy of the states vanish. The index $n$ runs from 0 (ground states) to 7
(seventh excited states), while the column with $n=\infty$ gives the analytical value
obtained for $\xi=\infty$ and on resonance where $a=\infty$. The latter is
$e^{2\pi/s_0}$ for the energy ratios and $e^{\pi/s_0}$ for the coherence length ratios.}
\label{tab}
\end{table}

\section{Discussion}
Using the Born-Oppenheimer approximation, we have studied both
qualitatively and quantitatively, the three-body bound states 
of two heavy impurities in a Bose condensate of light bosonic 
particles where the impurity-boson interaction is close to 
resonance. This correspond to large impurity-boson scattering length
and opens up the study of Efimov three-body states and their scaling
properties in the presence of a background medium of condensed 
bosons. A previous study \cite{zinner2013} argued that this implies a competition 
between the impurity-boson scattering length and the condensate
coherence length, and that the smallest of those scales will 
determine the estimated number of Efimov three-body states to 
expect. In fact, when the impurity-boson scattering length 
diverges, the condensate coherence length is the sole scale 
left to determine this number and it becomes an effective 
large distance cut-off on the attractive inverse-square
potential that is the origin of the Efimov effect.

We solved the bound state problem of the two impurities
interacting via the Born-Oppenheimer potential obtained 
by integrating out the light bosonic particle that was 
assumed to have a non-trivial dispersion due to the 
presence of the condensate (with a linear momentum 
dependence at small momenta). This allowed us to study
the flow of the bound state energies with condensate
coherence length. We find that once the energy scale
given by the coherence length becomes of order the 
three-body binding energy without any condensate, then 
the three-body state becomes strongly modified and eventually
gets pushed into the continuum. This is similar to the 
behaviour seen for Efimov states in the presence of a 
degenerate Fermi sea \cite{macneill2011,nygaard2011}. In fact, 
one can map the behaviours by changing Fermi energy, $E_F$, to 
the coherence length energy scale, $E_\xi=\hbar^2/M\xi^2$, where
$\xi$ is the coherence length. However, one needs to notice that
$M$ is the impurity mass. While we have assumed that $M$ is much larger
than the mass of the degenerate bosons, we expect that a similar
kind of scaling can be found for other mass ratios. This would 
require a formalism more general than the Born-Oppenheimer 
approach used here and in Ref.~\cite{macneill2011}. Furthermore,
it would be interesting to study the influence of fluctuations 
in the Bose condensate in similar fashion to what was discussed
in Ref.~\cite{macneill2011} for a Fermi sea where fluctuations 
in the form of particle-hole pairs have been considered.

In order to experimentally address the physics discussed here, a mass-imbalanced
system is needed. As noted above, we do expect similar behaviour also for mass-balanced
cases but this is not the topic of the present discussion. To achieve large
mass ratios some recent experiments with Helium are of interest \cite{borbely2012,knoop2012a} and in
particular the possibility to pursue mixtures of $^{4}$He and $^{87}$Rb \cite{knoop2012b}. 
Other mixtures of interest are  
$^7$Li and $^{133}$Cs \cite{mudrich2002}, $^7$Li and $^{87}$Rb \cite{marzok2009}, 
or mixtures of Ytterbium isotopes (mass numbers
168-176) \cite{onomoto2007,kitagawa2008} with $^7$Li or $^4$He. Estimates 
of the effects on the number of bound states the two former cases
have been given in Ref.~\cite{zinner2013}. For smaller mass imbalances we
mention $^{23}$Na with $^{40}$K \cite{wu2012} and $^{87}$Rb
with $^{133}$Cs \cite{takekoshi2012,weber2010,spethmann2011} as promising
experimentally studied systems. This requires theoretical ventures beyond
the Born-Oppenheimer approximation which we will pursue in the future. 
A key issue is experimental control over both the condensate coherence
length, $\xi$, and the impurity-boson scattering length, $a$. The most
common way to tune such parameters is through magnetically tunable 
Feshbach resonances but also optical Feshbach resonances have been 
studied \cite{chin2010}. One could then imagine an optical tuning of 
one of these parameters and a magnetic tuning of the other. Another 
possibility is to use the fact that $\xi=1/\sqrt{8\pi n_0 a_B}$ depend on not only the 
boson-boson scattering length, $a_B$, but also on the condensate density, $n_0$.
Changing $n_0$ could therefore accomplish similar things.
 
There is great experimental interest in strongly interactin Bose gases at the moment
both in three \cite{papp2008,pollack2009,navon2011,wild2012,makotyn2014} 
and in two dimensional systems \cite{na2013,fletcher2013,makhalov2014}.
Recent theoretical works addressing such experiments \cite{sykes2014,hudson2014} have shown
the importance of the few-body states (two- and three-body) present in the strongly interacting
gases. In similar experiments with mass-imbalanced mixtures of bosons, the few-body
states should play an equally important role. As we have discussed, the number
of bound states to be expected depends on the intra-species interaction strength and
the results presented here can guide one in estimating the bound state spectrum.

The main feature of our setup is the linear dispersion at 
low energy of the bosons. We obtain this through the presence
of a condensate background medium. However, it is easy to imagine
other ways to achieve the same thing. Systems like Graphene \cite{neto2009} 
and topological insulator surfaces \cite{hasan2010,qi2011} 
are examples of linear dispersions at low energy. The influence
of the electrons in these systems on impurities is a topic 
of great interest in the solid-state and condensed-matter 
communities \cite{balatsky2006}. More generally, we may ask 
what happens with modified dispersion and three-body systems, and
in different dimensionalities. For instance, applying spin-orbit coupling
to atomic systems \cite{zhai2012} has recently been shown to provide an 
interesting Efimov spectrum due to a modification of the dispersion \cite{shi2013}.
The recently discovered Efimov effect for spin-polarized fermions in 
two dimensions \cite{nishida2013} and the presence of deeply bound borromean 
states \cite{artem2013} implies that such a system could have a 
transition of its Fermi surface and thus of the dispersion. These 
are very interesting cases in which to study few-body physics in a 
many-body background.

\appendix

\section{Integrals}\label{appA}
The first integral that appears in Eq.~\eqref{central} can be solved exactly and yields
\begin{align}
&\int_{0}^{\infty} dx \left[\frac{x^2}{[x^4+x^2/(\alpha\xi)^2]^{1/2}+1}-1\right]=
-\frac{2}{\pi}\frac{1}{\alpha\xi}\left[-1\right.& \nonumber\\
&\left.+\frac{(\alpha\xi)^2}{\sqrt{2+8(\alpha\xi)^4}}\sqrt{\sqrt{1+4(\alpha\xi)^4}+1}\right.&\nonumber\\
&\left.\times\left(\textrm{ArcTan}\left[\sqrt{\frac{2}{\sqrt{1+4(\alpha\xi)^4}-1}}\right]-\pi\right)\right.&\nonumber\\
%&\phantom{-\frac{2}{\pi}\frac{1}{\xi}[ -1} 
&\left.+\frac{(\alpha\xi)^2}{\sqrt{2+8(\alpha\xi)^4}}\sqrt{\sqrt{1+4(\alpha\xi)^4}-1}\right.&\nonumber\\
&\left.\times\textrm{ArcTanh}\left[\sqrt{\frac{2}{\sqrt{1+4(\alpha\xi)^4}+1}}\right] \right].&
\end{align}
To first order in $A^2=\tfrac{1}{(\alpha\xi)^2}$ we obtain the simple expression
\begin{align}
-\frac{2}{\pi}\left(1+\frac{1}{4}A^2\right).\label{kinterm}
\end{align}
In the opposite limit where $A^2\gg 1$, the integral becomes 
\begin{equation}
-A-\tfrac{\pi}{2}\tfrac{1}{A},\label{kinasymp}
\end{equation}
to leading order in $A$.

The second integral 
\begin{align}
\int_{0}^{\infty} dx \frac{x \sin(\alpha R x)}{[x^4+A^2x^2]^{1/2}+1}\label{sinint}
\end{align}
is significantly more complicated due to the oscillatory factor. Numerically it 
is one of the most difficult tasks to compute highly oscillatory functions. Fortunately, 
there are limits in which analytical results can still be obtained. This is 
the case for the important limit of $A$ small. To second order in $A^2$ we have
\begin{align}
&\int_{0}^{\infty} dx \frac{x \sin(\alpha R x)}{x^4+(1+A^2/2)x^2-A^4/8}=&\nonumber\\
&\frac{\pi}{2H}\left[F^2\,\textrm{exp}(-\alpha R F)-G^2\,\textrm{cos}(\alpha R G)\right],&
\end{align}
where
\begin{align}
&F=\frac{2+A^2+\sqrt{4+4A^2+3A^4}}{4},\,\,\,G=F-1-\frac{A^2}{2}&
\end{align}
and $H=F^2+G^2$.
To second order this reduces to the simple result 
\begin{align}
\frac{\pi}{2}\textrm{exp}(-\alpha R-\frac{1}{4}\alpha R A^2).\label{expterm}
\end{align}

Using the results of Eqs.~\ref{kinterm} and \ref{expterm} in Eq.~\eqref{central} we obtain the eigenvalue equation for $\alpha$
\begin{align}
\alpha R + \frac{1}{4} \frac{R}{\alpha\xi^2}=\frac{R}{a}+\textrm{exp}(-\alpha R-\frac{1}{4}\frac{R}{\alpha\xi^2}),\label{2order}
\end{align}
which is our result for general scattering lengths. Solutions must still be consistent with the approximation $A^2\ll 1$ that 
was used to arrive at this result. We have checked this {\it a posteriori} and find that a sufficient condition is $R\ll\xi$. 
However, this is not a necessary condition as can be seen in the following way. Consider the limit of large $R$ in
Eq.~\eqref{2order} defined as the limit in which we drop the exponential term and find that Eq.~\eqref{2order} reduces
to
\begin{equation}
\frac{1}{a}=\alpha+\frac{1}{4\alpha\xi^2},\label{2twobody}
\end{equation}
We thus see that the two-body problem is recovered in this limit. Note that this result is 
consistent with the approximation $A^2\ll 1$ for large $\xi$ (which is a basic assumption on the formalism as discussed in 
Sec.~\ref{theory}).

The regime of $A^2\gg 1$ is more difficult. We may consider the extreme limit where we approximate the integrand in 
Eq.~\eqref{sinint} to obtain
\begin{align}
&\int_{0}^{\infty} dx \frac{x \sin(\alpha R x)}{Ax+1}=&\nonumber\\
&\frac{\frac{2 A}{\alpha R}-2\textrm{Ci}
\left[\frac{\alpha R}{A}\right]\sin\left(\frac{\alpha R}{A}\right)-
\cos\left(\frac{\alpha R}{A}\right)(2\textrm{Si}\left[\frac{\alpha R}{A}\right]-\pi)}{\sqrt{2\pi}A^2},&
\end{align}
which is the sine transform of the function $x/(Ax+1)$. Here $\textrm{Si}(x)$ and $\textrm{Ci}(x)$ are the Sine and Cosine integrals. 
To be more precise one should split the integral in Eq.~\eqref{sinint}
according to 
values of $x^2<A^2$ and $x^2>A^2$. However, the presence of oscillatory factor $\sin(\alpha R x)$ implies that the latter part will 
give a very small contribution for $A^2\gg 1$ and we neglect it. The argument of the above expansion is 
$\alpha R/A=\tfrac{R}{\xi}\tfrac{1}{(\alpha\xi)^2}$. If we assume that $R\sim\xi$, this is a large number ($A^2\gg 1$) and we
can expand the result above to obtain
\begin{align}
\int_{0}^{\infty} dx \frac{x \sin(\alpha R x)}{Ax+1}\to \frac{\xi}{R}-\frac{1}{2}\pi^2 (\alpha\xi)^2\label{sinasymp}
\end{align}
to second order in $\alpha\xi$ which is assumed small.

\section{Impurity Potential at Unitarity}\label{appB}
We now assume that $a$ diverges and thus neglect the left-hand side of Eq.~\eqref{central}. Furthermore,
we assume that $A^2=\tfrac{1}{(\alpha\xi)^2}\ll 1$ and thus truncate the integrals. Eq.~\eqref{central} now reduces to the 
simple form
\begin{equation}
\alpha R+\frac{1}{4}\alpha R A^2 = \textrm{exp}(-\alpha R-\frac{1}{4}\alpha R A^2).\label{eigeq}
\end{equation}
This is readily solved since we have $x_0=\textrm{exp}(-x_0)$ is solved by $x_0=0.567$ and we thus have 
$\alpha R+\tfrac{1}{4}\alpha R A^2=x_0$, which is an implicit second degree equation for $\alpha$. The 
solutions are
\begin{align}
\alpha R=\frac{x_0\pm\sqrt{x_{0}^{2}-R^2/\xi^2}}{2}.\label{roots}
\end{align}
The existance of a real solution is thus seen to require $R/\xi\leq x_0$. In addition, we see that 
$\alpha R=\tfrac{x_0}{2}$ for $R=x_0\xi$. However, we need to consider also the condition $\alpha \xi\ll 1$
($A^2\gg 1$).
which was a prerequisite for obtaining the eigenvalue equation in the form of Eq.~\eqref{eigeq}. By expansion
around $R/\xi\to 0$ of the roots in Eq.~\eqref{roots}, one sees that the one with the minus sign goes to 
zero in this limit and must be discarded. 
The solution with the plus sign in Eq.~\eqref{roots} is shown in Fig.~2 of Ref.~\cite{zinner2013} as the solid (blue) 
curve. 
We have checked that for small $\alpha \xi$, a numerical solution of the integrals of Appendix~\ref{appA}
yields the same result.

A simple second order approximation around $R/\xi=0$, can be found from the acceptable solution with the
plus sign in Eq.~\eqref{roots}. Expansion yields
\begin{align}
\alpha R = x_0-\frac{1}{4x_0}\left(\frac{R}{\xi}\right)^2+\textrm{O}(R^4/\xi^4).
\end{align}
The second order result is shown as the dotted (black) line in Fig.~2 of Ref.~\cite{zinner2013}.

\section{Efimov Effect in the Born-Oppenheimer Limit}\label{appC}
For completeness of presentation and reference, in this appendix we discuss the Efimov 
effect in the heavy-heavy-light system in the 
absence of a condensate background felt by the light particle.
Following the Born-Oppenheimer description, this energy is now an effective potential for the two
heavy particles as function of their distance $R$. So we have the heavy-heavy Schr{\"o}dinger equation
\begin{equation}\label{seq}
\left[-\frac{\hbar^2}{M}\nabla^{2}_{\bm R}+E_R(\bm R)\right]\Phi(\bm R)=E\Phi(\bm R),
\end{equation}
where the kinetic energy is missing a factor of 2 in the denominator since the reduced mass in the 
heavy-heavy system is $M/2$. Here $E$ is the total energy of the system of light-heavy-heavy type.
This equation can be massaged into a nicer form by defining $u(\bm R)=R\Phi(\bm R)$, $\kappa^2=-ME/\hbar^2>0$
(we are looking for bound states with $E<0$), and $z=\kappa R$. This produces
\begin{equation}\label{ueq}
\frac{d^2u(z)}{dz^2}-u(z)+\frac{\beta(a,z)^2}{z^2}u(z)=0,
\end{equation}
where 
\begin{equation}
\beta(a,z)^2=\frac{x_{0}^{2}}{2}\frac{M}{m}\left[1+\frac{1}{x_{0}^{2}(1+e^{x_0})}\frac{R}{a}\right],
\end{equation}
is still a function of $z$. However, if we consider $z\ll \kappa a$ ($R\ll a$) or $|a|\to\infty$, we can
drop the second term and get a constant $\beta$. In this case, the solution to Eq.~\eqref{ueq} which has the 
correct boundary condition at $z\to\infty$ (exponentially decreasing, $e^{-z}$) contains a modified
Bessel of the second kind, $K_n(z)$. However, the order has to be imaginary, i.e. $n\to i(\beta^2-1/4)$, 
and we arrive at $u(z)=\sqrt{z}K_{i(\beta^2-1/4)}(z)$ (not normalized). In our case, $M\gg m$, so $\beta^2\gg 1$, i.e. 
it is always a positive imaginary order. The wave function $u(z)$ has the following behavior at small $z$
\begin{equation}\label{ulim}
u(z)\propto \sqrt{z}\cos\left(s_0 \textrm{log}(z)\right),
\end{equation}
where $s_0=\sqrt{\beta^2-1/4}$. This is the origin of the log-periodic behavior of the states found by Efimov
when $|a|\to\infty$. In order to show this, we observe an interesting scaling invariance of Eq.~\eqref{seq};
if we consider $\Phi(\lambda \bm R)$, then it solves the Schr{\"o}dinger equation with energy $\lambda^2 E$ when
$|a|\to\infty$ for $\lambda$ a real number. However, Eq.~\eqref{ulim} further contrains the choice of 
$\lambda$. Using $z=\kappa R$, we have 
\begin{equation}
\Phi(\bm R)\propto \sqrt{\kappa R}\cos\left(s_0 \textrm{log}(\frac{R}{R_0})\right),
\end{equation}
where $R_0$ is a necessary short-distance cut-off which comes from the repulsive cores of
atoms at short distance. In order to preserve the $\lambda$ scaling, we need to require
$s_0 \textrm{log}(\lambda)=n\pi$, where $n$ is an integer. We immediately see that there
is an infinitude of three-body bound states with energies
\begin{equation}
E_n=E_0 e^{-2\pi n/s_0},
\end{equation}
which is the famous Efimov effect. The ground state is related to $R_0$ since $E_0\sim -\hbar^2/2MR_{0}^{2}$ 
(up to constants of order 1 that we have neglected for simplicity). For $M\gg m$, $s_0$ will be large and 
will therefore in turn lead to a small scale factor $e^{-2\pi n/s_0}$, which means that the spectrum of 
three-body bound states is quite dense, there are many three-body bound states around! For fininte $a$, it
is possible to show that the above considerations are good up to $R\sim |a|$, and that the number of 
bound states is approximately $N\sim s_0\textrm{log}(|a|/R_0)/\pi$ and increases with $s_0$, i.e. with
$\sqrt{M/m}$.

\end{document}